\documentclass[10pt, twocolumn]{article}

\usepackage{geometry}
\usepackage{amsthm,amsmath}
\usepackage[square]{natbib}
\setcitestyle{numbers,comma}
\usepackage{latexsym, amsfonts, amssymb, amscd}
\usepackage{bbm}
\usepackage[T1]{fontenc}
\usepackage{graphicx}
\usepackage[utf8]{inputenc}
\usepackage{abstract}
\usepackage{tikz}

\usepackage[font={footnotesize}]{caption}

\usepackage{todonotes}

\usetikzlibrary{shapes.geometric}
\usetikzlibrary{positioning}

\usepackage{titlesec}

\titlespacing*{\section}
  {0pt}% retrait à gauche
  {0.5ex plus .2ex minus .2ex}% espace avant
  {0.5ex plus .2ex}% espace après
  [0pt]% retrait à droite

\makeatletter
\renewenvironment{thebibliography}[1]
     {\section*{\refname}%
      \@mkboth{\MakeUppercase\refname}{\MakeUppercase\refname}%
      \list{\@biblabel{\@arabic\c@enumiv}}%
           {\settowidth\labelwidth{\@biblabel{#1}}%
            \leftmargin\labelwidth
            \advance\leftmargin\labelsep
            \@openbib@code
            \usecounter{enumiv}%
            \let\p@enumiv\@empty
            \renewcommand\theenumiv{\@arabic\c@enumiv}}%
      \setlength{\itemsep}{0cm}
      \sloppy
      \clubpenalty4000
      \@clubpenalty \clubpenalty
      \widowpenalty4000%
      \sfcode`\.\@m}
     {\def\@noitemerr
       {\@latex@warning{Empty `thebibliography' environment}}%
      \endlist}
\makeatother

\usepackage{titling}
\pretitle{\begin{center}\LARGE}
\posttitle{\par\end{center}\vspace{0.5em}}
\preauthor{\begin{center}\small
           \begin{tabular}[t]{c}}
\postauthor{\end{tabular}\par\end{center}}
\renewcommand{\v}[1]{\boldsymbol{#1}}

\newcommand\blfootnote[1]{%
  \begingroup
  \renewcommand\thefootnote{}\footnote{#1}%
  \addtocounter{footnote}{-1}%
  \endgroup
}

\title{Spike-based probabilistic inference with correlated noise}

\date{12.03.2017}
\author{Ilja Bytschok, Dominik Dold, Johannes Schemmel, Karlheinz Meier and Mihai A. Petrovici\textsuperscript{*}}

\begin{document}

\maketitle

It has long been hypothesized that the trial-to-trial variability in neural activity patterns is not a mere byproduct of computation, but plays an important role in information processing.
A steadily increasing body of evidence suggests that this activity is a hallmark of ongoing probabilistic inference employed by the brain to interpret and respond to sensory input \cite{kording2004bayesian, fiser2010statistically, jezek2011theta, rich2016decoding}.

The neural sampling hypothesis \cite{buesing2011neural} interprets the stochastic activity of neurons as sampling from an underlying probability distribution.
This distribution can be shaped by sensory data to obtain an accurate probabilistic representation of the relevant data space \cite{leng2016spiking}.

The abstract model of neural sampling has been shown to be compatible with biologically observed dynamical regimes of spiking neurons \cite{petrovici2016stochastic}.
In these studies, high-frequency Poisson spike trains were used as a source of stochasticity.
While this representation of diffuse synaptic input has become commonplace in theoretical and computational neuroscience, it often discounts the fact that cortical neurons may share a significant portion of their presynaptic partners, thus receiving correlated noise.
This can have profound impact on the computation these neurons are required to perform.
Importantly, the issue of shared noise channels is not only relevant in biology, but can become particularly problematic in artificial implementations of neural networks \cite{furber2016large} which need to deal with limited bandwidth and therefore a limited amount of independent noise channels.

Here, we address this issue in the context of neural sampling.
We show that a sampled Boltzmann distribution over binary random variables is indeed modified by shared input correlations, but that it remains a Boltzmann distribution, albeit with different parameters.
Since the effects of shared input correlations can therefore be compensated by changing network parameters, a sampling network receiving correlated input can be trained using the same learning rules as for independent stochastic units, with no penalty to their performance in reproducing the target distribution.

For our studies we use the sampling framework established in \cite{petrovici2016stochastic}.
In this framework, the firing activity of a network with $N$ Leaky Integrate-and-Fire (LIF) neurons is represented by a vector of binary random variables, $\v z\in\{0,1\}^N$ (Fig.~\ref{fig:1}A).
When neurons enter a high-conductance state (due to, e.g., background synaptic bombardment), synaptic weights can be chosen such that the network approximately samples from a Boltzmann distribution
\begin{equation}
    p(\v z) \propto \exp\left( \v z^T \v W \v z + \v z^T \v b \right)
    \label{eqn:boltzmann}
\end{equation}
defined by an abstract parameter set $(\v W, \v b)$, which represent the connection matrix and bias vector, respectively (Fig.~\ref{fig:1}B).

The above equation implicitly assumes that the only correlations between states arise from the connection matrix $\v W$.
This is no longer true when noise is correlated, as shared noise induces synchronous firing and silent states.
To illustrate this point, we show the effect on a distribution sampled from a neuron pair with $\v W = \v 0$ and $\v b = \v 0$ (Fig.~\ref{fig:1}C), where correlated noise shifts probability mass from mixed states $(0,1)$ and $(1, 0)$ into synchronous states $(0,0)$ and $(1,1)$.
This differs from simply increasing $W_{12}$, as a weight increase only induces more synchronous firing (Fig.~\ref{fig:1}D).
In this case, probability mass shifts from $(0, 0)$, $(1,0)$ and $(0, 1)$ towards $(1,1)$ since $W_{ij}$ only affects $(1,1)$ pairs in Eqn.~\ref{eqn:boltzmann}.
In this two-neuron example, this effect can be easily counteracted by modifying $\v b$ in addition to $\v W$, but it is not obvious whether and how this can be applied to larger networks.

The relationship between shared noise and parameters $(\v W, \v b)$ becomes more intuitive when changing the state space from $\v z\in\{0,1\}^N$ to {$\v z'\in \{-1, 1\}^N$}, while preserving the state distribution.
In this state space, an increase in $W_{12}$ affects all states (Fig.~\ref{fig:1}E):
The probabilities of correlated states $(-1, -1)$ and $(1, 1)$ increase, while the probabilities of uncorrelated states $(-1,1)$ and $(1, -1)$ decrease.
This mimics the effect of correlated noise (Fig.~\ref{fig:1}C), allowing us to find a bijective mapping
\begin{equation}
    \Delta W' = f(s), \quad \text{with} \; f := g^{-1} \circ h
    \label{eqn:rmap}
\end{equation}
between the shared noise ratio $s$ and weight change $\Delta W'$ in the $\{-1, 1\}^N$ space (Fig.~\ref{fig:1}F,G) using the correlation coefficient $r$ as an intermediate ($h$ and $g$ are the maps from $s$ and $\Delta W'$ to $r$, respectively).

However, the state space $\{-1, 1\}^N$ appears incompatible with neuron states defined by spikes, as a neuron would need to excite a postsynaptic partner when it spikes and inhibit it otherwise.
Nevertheless, it is possible to find a mapping between the network parameters in the state spaces $z\in\{0,1\}^N$ and $z'\in\{-1, 1\}^N$ (using $z = (z'+1)/2$) that conserves probabilities:
\begin{align}
    \v W = 4 \v W' \, ,\quad \v b = 2 \v b' + 2\v W'\quad.
    \label{eqn:wbmap}
\end{align}
With these two maps (Eqn.~\ref{eqn:rmap} and \ref{eqn:wbmap}), it is now possible to compensate for shared noise by an appropriate parameter modification in a spiking network, whose states reside naturally in the $\{0,1\}^N$ space.
The required sequence of steps and the results of this compensation are shown in Fig.~\ref{fig:1}F-H.
Although some discrepancies between the target and sampled distributions remain, due to the imperfect translation between weights in the abstract model (Eqn. \ref{eqn:boltzmann}) and synaptic weights in an LIF network, the compensation of correlated noise can be seen to perform well.

Disregarding the precise nature of the compensation, the mere fact that such a mapping exists demonstrates that correlated noise does not cause the sampled distribution to leave the Boltzmann family.
This is essential for the ability of such networks to learn from data.
For an untrained network, the existence of noise correlations simply shifts its initial position in the parameter space, leaving it amenable to the same learning rules that can be used in the correlation-free case.
In other words, because in the $\{-1,1\}$-coding the correlated noise can be compensated according to Eq. 2 by some $\Delta W'$, and because the $\{-1,1\}$-coding can be transformed into a $\{0,1\}$-coding while keeping the state probabilities invariant, a learning rule for Boltzmann machines will also find that distribution in the $\{0,1\}$-coding.
One such learning rule is the wake-sleep algorithm, which is Hebbian in nature and iteratively shifts the network activity such that it best reproduces the chosen set of training data.
For networks whose states live in the $\{0,1\}^N$ space, this learning rule can be written as
\begin{align}
  \delta W_{ij} &\propto p_{ij}^{\text{data}}(1,1) - p_{ij}^{\text{net}}(1,1) \quad , \\
  \delta b_i &\propto p_i^{\text{data}}(1) - p_i^{\text{net}}(1) \quad ,
\end{align}
where $p_{ij}(1,1)$ is shorthand for $p(z_i=1,z_j=1)$.
Fig.~\ref{fig:1}I shows the evolution of a sampling network of LIF neurons receiving correlated noise that is trained with the above learning rule.
Note how the end result is even superior to the compensation technique described above, as the learning rule is applied directly to the LIF network and is therefore agnostic about discrepancies between LIF networks and Boltzmann machines.

In conclusion, we have shown that spiking networks performing neural sampling are impervious to noise correlations when appropriately trained.
Therefore, if such computation happens in cortex, network plasticity does not need to take particular account of shared noise inputs.
For an artificial embedding of such networks, our compensation technique allows a straightforward transfer between platforms with different architecture and bandwidth constraints.

\usetikzlibrary{arrows}
\definecolor{darkred}{rgb}{0.55,0.,0.}
\definecolor{darkblue}{rgb}{0.,0.,0.55}
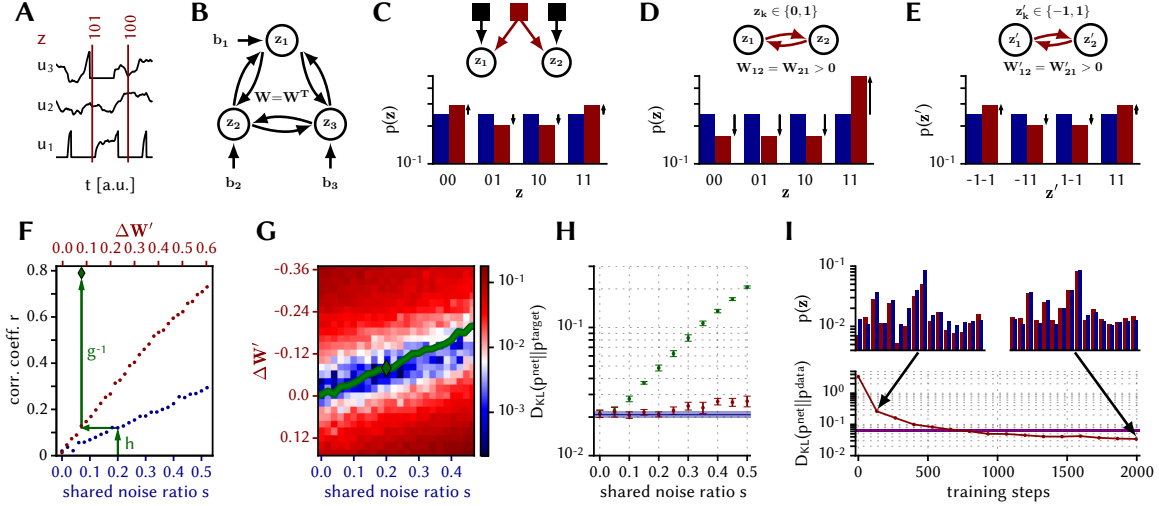
\begin{figure*}[t]
    \centering
    \begin{tikzpicture}[
            font=\sffamily,
            line width=1pt,
            scale=1.0,
            ->,
            shorten >=1pt,
            shorten <=1pt,
            >=latex,
            transform shape,
            square/.style={regular polygon,regular polygon sides=4},
        ]
        %~ \begin{scope}[shift={(-4.2,1.9)},scale=.65]
            %~ \node[draw, circle, minimum size=2.2cm] (m) at (0,0) {};
            %~ \node[minimum size=0pt] (b) at (0,-1.7) {$b$};
            %~ \foreach \n in {0,...,2}
            %~ {
                %~ \node[draw, circle, minimum size=0.6cm] (c\n) at (120*\n+90:.6) {};
                %~ \draw (b.north-|c\n) to (c\n.south);
            %~ }
            %~ \path (m) edge [out=30, in=80, looseness=3.5, <->] node[above right] {$\bs w$} (m);
        %~ \end{scope}
        \begin{scope}[shift={(-4.2,2.2)},scale=.55]
            \foreach \n [evaluate=\n as \i using \n-1]  in {1,...,3}
            {
                \node[draw, circle, minimum size=0.4cm] (c\n) at (120*\i+90:1.35) {$\bf z_\n$};
            }
            \node (c0l) at (110:1.35) {};
            \node (c0r) at (70:1.35) {};
            \node (text) at (0:0) {$\bf W\hspace{-1mm} =\hspace{-1mm} W^T$};

			\node[left = .75cm of c1] (b1) {$\bf b_1$};
			\draw[->] (b1.east) -- (c1.west);
			\node[below = .75cm of c2] (b2) {$\bf b_2$};
			\draw[->] (b2.north) -- (c2.south);
			\node[below = .75cm of c3] (b3) {$\bf b_3$};
			\draw[->] (b3.north) -- (c3.south);

			\path (c0l.south) edge [bend left=20, ->] node [right] {} (c2.north);
            \path (c2.north) edge [bend right=-20, ->] node [left] {} (c0l.south);
            \path (c0r.south) edge [bend right=20, ->] node [left] {} (c3.north);
            \path (c3.north) edge [bend left=-20, ->] node [right] {} (c0r.south);
            \path (c2.east) edge [bend right=20, ->] node [left] {} (c3.west);
            \path (c3.west) edge [bend left=-20, ->] node [right] {} (c2.east);

        \end{scope}
        \pgfresetboundingbox
        \draw[use as bounding box,inner sep=0pt] node {\includegraphics[width=\textwidth]{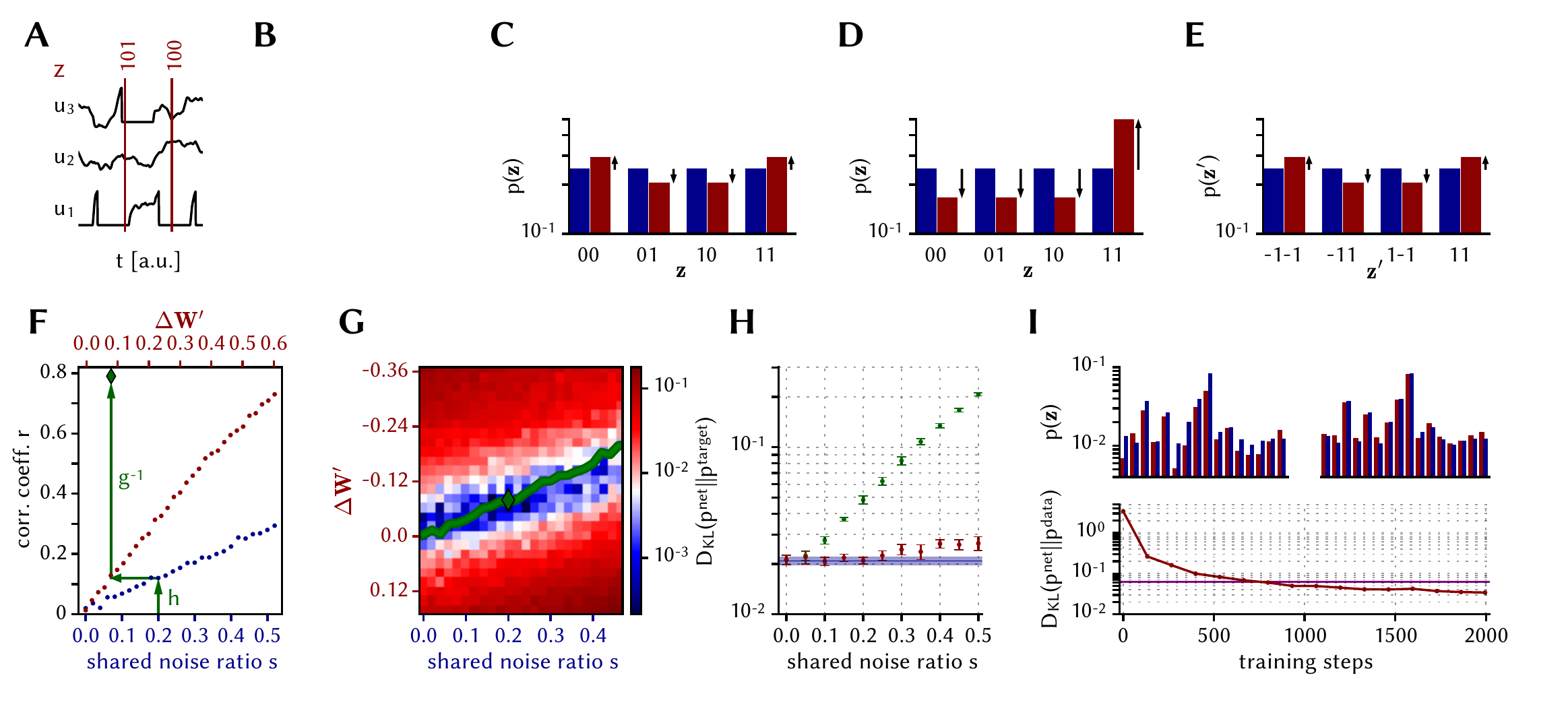}};
        %~ \begin{scope}[shift={(-.65,2.3)},scale=.5]
            %~ \node[draw, circle, minimum size=.65cm] (a) at (1.65,1.25) {$\bf z_1$};
            %~ \node[draw, circle, minimum size=.65cm] (b) at (1.65,0.15) {$\bf z_2$};
            %~ \node[fill, square, minimum size=.65cm] (c) at (0,1.25) {};
            %~ \node[fill, square, minimum size=.65cm] (d) at (0,0.15) {};
            %~ \draw[->] (c.east) +(-0.15em,0) coordinate (b1) -- (a.west |- b1);
            %~ \draw[->] (d.east) +(-0.15em,0) coordinate (b1) -- (b.west |- b1);
        %~ \end{scope}
        %~ \begin{scope}[shift={(-.55,2.7)},scale=.5]
            %~ \node[draw, circle, minimum size=.65cm] (a_1) at (1.65,1.4) {$\bf z_1$};
            %~ \node[draw, circle, minimum size=.65cm] (b_1) at (1.65,0) {$\bf z_2$};
            %~ \node (anode) at (1.45, 1.3) {};
            %~ \node (bnode) at (1.45, 0.1) {};
            %~ \node[fill, square, minimum size=.65cm] (c_1) at (0,1.4) {};
            %~ \node[fill, square, minimum size=.65cm] (d_1) at (0,0) {};
            %~ \node[fill, darkred, square, minimum size=.65cm] (e) at (0,0.7) {};
            %~ \draw[->] (c_1.east) +(-0.15em,0) coordinate (b1) -- (a_1.west |- b1);
            %~ \draw[->] (d_1.east) +(-0.15em,0) coordinate (b1) -- (b_1.west |- b1);
            %~ \draw[darkred, ->] (e.east) +(-0.15em,0) -- (anode.west);
            %~ \draw[darkred, ->] (e.east) +(-0.15em,0) -- (bnode.west);
        %~ \end{scope}
        \begin{scope}[shift={(-1.55,2.65)},scale=.5]
            \node[draw, circle, minimum size=.65cm] (a_1) at (0.,0) {$\bf z_1$};
            \node[draw, circle, minimum size=.65cm] (b_1) at (2.,0) {$\bf z_2$};
            \node (anode) at (0.3, 0.1) {};
            \node (bnode) at (1.7, 0.1) {};
            \node[fill, square, minimum size=.65cm] (c_1) at (2.,1.35) {};
            \node[fill, square, minimum size=.65cm] (d_1) at (0,1.35) {};
            \node[fill, darkred, square, minimum size=.65cm] (e) at (1.,1.35) {};
            \draw[->] (d_1.south) +(0,.5em) -- (a_1.north);
            \draw[->] (c_1.south) +(0,.5em) -- (b_1.north);
            \draw[darkred, ->] (e.south) +(0,0.25em) -- (anode.north);
            \draw[darkred, ->] (e.south) +(0,0.25em) -- (bnode.north);
        \end{scope}
        \begin{scope}[shift={(2.,2.95)},scale=.5]
			\node (text_state) at (1.0,0.75) {$\bf z_k \in \{0,1\}$};
			\node (w_state) at (1.0,-0.75) {$\bf W_{12} = W_{21} > 0$};
            \node[draw, circle, minimum size=.65cm] (a) at (0,0) {$\bf z_1$};
            \node[draw, circle, minimum size=.65cm] (b) at (2.0,0) {$\bf z_2$};
            \path (a.east) edge [bend left=20, darkred, ->] node [right] {} (b.west);
            \path (b.west) edge [bend right=-20, darkred, ->] node [left] {} (a.east);
        \end{scope}
         \begin{scope}[shift={(5.55,2.95)},scale=.5]
			\node (text_state) at (1.0,0.75) {$\bf z'_k \in \{-1, 1\}$};
			\node (w_state) at (1.0,-0.75) {$\bf W'_{12} = W'_{21} > 0$};
            \node[draw, circle, minimum size=.65cm] (a) at (0,0) {$\bf z'_1$};
            \node[draw, circle, minimum size=.65cm] (b) at (2.0,0) {$\bf z'_2$};
            \path (a.east) edge [bend left=20, darkred, ->] node [right] {} (b.west);
            \path (b.west) edge [bend right=-20, darkred, ->] node [left] {} (a.east);
        \end{scope}
         %~ \begin{scope}[shift={(7.2,-.5)},scale=.1]
			%~ \node[fill, square, minimum size=2.cm] (b_1) at (7., -1.5) {};
			%~ \node[fill, square, minimum size=2.cm] (b_2) at (7., -7.5) {};
			%~ \node[fill, square, minimum size=2.cm] (b_3) at (7., -16.5) {};
%~
            %~ \node[draw, circle, minimum size=2.cm] (a) at (2.,0.) {};
            %~ \node[draw, circle, minimum size=2.cm] (b) at (2.0,-3.) {};
            %~ \node[draw, circle, minimum size=2.cm] (c) at (2.,-6.) {};
            %~ \node[draw, circle, minimum size=2.cm] (d) at (2.0,-9.) {};
            %~ \node[draw, fill, circle, radius=1pt] (e) at (2.,-11) {};
            %~ \node[draw, fill, circle, radius=1pt] (f) at (2.0,-12) {};
            %~ \node[draw, fill, circle, radius=1pt] (g) at (2.0,-13) {};
            %~ \node[draw, circle, minimum size=2.cm] (h) at (2.,-15.) {};
            %~ \node[draw, circle, minimum size=2.cm] (i) at (2.0,-18.) {};
%~
            %~ \draw[-stealth] (b_1.west)  -- (a.east);
			%~ \draw[-stealth] (b_1.west)  -- (b.east);
			%~ \draw[-stealth] (b_2.west)  -- (c.east);
			%~ \draw[-stealth] (b_2.west)  -- (d.east);
			%~ \draw[-stealth] (b_3.west)  -- (h.east);
			%~ \draw[-stealth] (b_3.west)  -- (i.east);
%~
			%~ \begin{scope}[scale = 5.]
			%~ \node (weight) at (-.4, -1.85) {$\bf \Huge{W}$};
			%~ \node (weight_east) at (-.35, -1.85) {};
			%~ \end{scope}
%~
			%~ \path (a.west) edge [bend right=20, -]  (weight.north);
			%~ \path (b.west) edge [bend right=20, -]  (weight.north);
			%~ \path (c.west) edge [bend right=20, -]  (weight.north);
			%~ \path (d.west) edge [-] (weight_east.east);
			%~ \path (h.west) edge [bend left=20, -]  (weight.south);
			%~ \path (i.west) edge [bend left=20, -]  (weight.south);
        %~ \end{scope}
        \begin{scope}
        \draw[->] (4.278,-1.18) -- (3.69,-2.);
        \draw[->] (6.32,-1.18) -- (7.18,-2.36);
        \end{scope}
        %~ \node[inner sep=0pt] (indep) at (3.05,2.65)
			%~ {\includegraphics[width=.08\textwidth]{fig/dummy_shared}};
		%~ \node[inner sep=0pt] (shared) at (3.05,1.4)
			%~ {\includegraphics[width=.08\textwidth]{fig/dummy_shared}};
		%~ \node[inner sep=0pt] (lif) at (6.6,2.65)
			%~ {\includegraphics[width=.07\textwidth]{fig/dummy_lif}};
		%~ \node[inner sep=0pt] (ising) at (6.6,1.4)
			%~ {\includegraphics[width=.07\textwidth]{fig/dummy_ising}};
    \end{tikzpicture}
    \caption{
        \textbf{(A)} Interpretation of neural dynamics as sampling in the state space $\{0,1\}^3$.
            Following a spike, a neuron enters a refractory period which is identified with a state $z=1$.
        \textbf{(B)} Exemplary architecture of a network with 3 neurons that samples from a Boltzmann distribution with parameters $\v W$ and $\v b$.
            In order to achieve the required stochastic regime, each neuron receives external noise in the form of Poisson spike trains (not shown).
        \textbf{(C)-(E)} Exemplary sampled distributions for a network of two neurons.
            The ``default'' case is the one where all weights and biases are set to zero (uniform distribution, blue bars).
        \textbf{(C)} Shared noise sources have a correlating effect, shifting probability mass into the (1,1) and (0,0) states (red bars).
        \textbf{(D)} In the $\{0,1\}^2$ space, increased weights introduce a positive shift of probability mass from all other states towards the (1,1) state (red bars), which is markedly different from the effect of correlated noise.
        \textbf{(E)} In the $\{-1,1\}^2$ space, increased weights have the same effect as correlated noise (red bars).
        \textbf{(F)} Dependence of the correlation coefficient $r$ between the states of two neurons on the change in synaptic weight $\Delta W'$ (red) and the shared noise ratio $s$ (blue).
            These define bijective functions $g$ and $h$ that can be used to compute the weight change needed to compensate the effect of correlated noise (Eqn.~\ref{eqn:rmap}) in the $\{-1,1\}^N$ space.
        \textbf{(G)} Study of the optimal compensation rule in a network with two LIF neurons.
            For simplicity, the ordinate represents weight changes for a network with states in the $\{-1,1\}^2$ space, which are then translated with Eqn.~\ref{eqn:wbmap} to corresponding parameters ($\v W, \v b$) for the $\{0,1\}^2$ state space.
            The colormap shows the difference between the sampled and the target distribution measured by the Kullback-Leibler divergence $\mathrm{D}_\mathrm{KL}(p^{\text{net}} || p^{\text{target}})$.
            The mapping provided by the compensation rule from Eqn.~\ref{eqn:rmap} is depicted by the green curve.
            Note that the compensation rule provides a nearly optimal parameter translation.
            Remaining deviations are due to differences between LIF dynamics and abstract Boltzmann machines.
        \textbf{(H)} Compensation of noise correlations in a network with two LIF neurons.
            The results are depicted for a set of 10 randomly drawn Boltzmann distributions over $\v z \in \{0,1\}^{10}$ (error bars).
            For a set of randomly chosen Boltzmann distributions, a ten-neuron network performs sampling in the presence of pairwise shared noise ratios $s$ (x-axis).
            The blue line marks the optimal sampling quality at $s=0$.
            For an increasing shared noise ratio, uncompensated noise (green) induces a significant increase in sampling error.
            After compensation following Eqn.~\ref{eqn:rmap} and \ref{eqn:wbmap}, the sampling quality is nearly completely restored.
            As before, remaining deviations are due to differences between LIF dynamics and abstract Boltzmann machines.
        \textbf{(I)} A 10-LIF-neuron network with shared noise sources ($s=0.3$ for each neuron pair) is trained with data samples generated from a target Boltzmann distribution (blue bars).
            During training, the sampled distribution becomes an increasingly better approximation of the target distribution (red line).
            For comparison, we also show the distribution sampled by an LIF network with parameters translated directly from the Boltzmann parameters following \cite{petrovici2016stochastic} (purple).
            The learning algorithm is able to outperform the direct-to-LIF translation because it is impervious to the differences between LIF dynamics and those of a Boltzmann machine.
            }
    \label{fig:1}
\end{figure*}
\blfootnote{\textsuperscript{*} Authors Bytschok, Dold and Petrovici contributed equally to this work. This research was supported by EU grants \#269921 (BrainScaleS), \#604102 (Human Brain Project) and the Manfred St\"ark Foundation.}

%\bibliographystyle{abbrv}
%\bibliography{bib}
\footnotesize
\bibliographystyle{alpha}

\vspace{7pt}

\end{document}